\RequirePackage{fix-cm}
\documentclass[twocolumn]{svjour3}          
\smartqed  
\usepackage{graphicx}
\journalname{Journal of Radioanalytical Nuclear Chemistry}
\begin{document}

\title{Activation cross-sections of longer lived products of proton induced nuclear reactions on cobalt up to 70 MeV
}


\author{F. Ditr\'oi  \and F. T\'ark\'anyi \and S. Tak\'acs \and A. Hermanne \and H. Yamazaki \and M. Baba \and A. Mohammadi
}


\institute{F. Ditr\'oi F. T\'ark\'anyi \and S. Tak\'acs\at
              Institute of Nuclear Research of the Hungarian Academy of Sciences \\
              Tel.: +36-52-509251\\
              Fax: +36-52-416181\\
              \email{ditroi@atomki.hu}           
           \and
           A. Hermanne \at
              Cyclotron Laboratory, Vrije Universiteit Brussel (VUB), Brussels, Belgium
           \and
              H. Yamazaki \and M. Baba \and A. Mohammadi \at
            Cyclotron Radioisotope Center (CYRIC) Tohoku University, Sendai, Japan  
}

\date{Submitted: 2013-1-15 / Accepted: }

\maketitle

\begin{abstract}
As a part of our series of studies on charged particle induced reactions on various target materials, proton induced excitation functions on natural cobalt have been measured by using stacked-foil technique. In these measurements $^{51}$Cr, $^{55,56,57,58}$Co, $^{51}$Cr, $^{52,54,56}$Mn and $^{56,57}$Ni radioisotopes have been identified. For the above isotopes the excitation functions were determined and compared with the literature data and with the results of EMPIRE and TALYS calculations taken from the TENDL 2011 library. The agreement with previous measurements was acceptable and we could also determine new cross-section data.
\end{abstract}

\section{Introduction}
\label{intro}
Cobalt is an element of the iron group, which is mainly used in production of alloys of good magnetic and corrosion properties as well as high durability. The radioactive cobalt isotopes are applied as high-intensity  $\gamma$-ray sources as well as tracers and references radioactive sources.  In addition to human oncological treatments $^{60}$Co is also used for sterilization of different products as food, medical tools and supplies, etc. Such high intensity  $\gamma$-radiation can annihilate pathogenic species. Unfortunately $^{60}$Co cannot be produced by proton activation on Co target.
Other radioactive cobalt isotopes are the $^{58}$Co, $^{57}$Co and $^{56}$Co. $^{57}$Co has shorter half-life than $^{60}$Co (272 d and 5.27 y respectively), so it can be effectively used in medical research. Because its low energy  $\gamma$-rays $^{57}$Co is used for clinical nuclear medical research, to calibrate SPECTs (Single Photon Emission Tomographs) and for use as a M\"ossbauer radioisotope \cite{7}. The application area of $^{58}$Co is the radiobiology, while $^{57}$Co and $^{56}$Co are intensively used in scientific and industrial research and practice. $^{57}$Co is used because of its low energy  $\gamma$-lines, while the advantages of $^{56}$Co are the strong high-energy  $\gamma$-lines. The radioisotope $^{55}$Co has short half-life, that's why it is extensively used in everyday medicine and research. Because the target element ($^{59}$Co) is monoisotopic, it is an ideal candidate for verification and improvement of theoretical nuclear reaction codes. $^{56,57,58}$Co are the radioisotopes, which are mainly used for industrial wear measurements by TLA (Thin Layer Activation) both produced from iron and from cobalt content of the alloy in question. If the alloy does not contain Fe or only Co, the wear curves, calculated from the measured cross-sections and  provided for the users makes possible to perform wear measurements also on these alloys.
This work is a part of our systematic study on light ion induced nuclear reactions on various elements for medical and industrial purposes, as well as for supporting theoretical calculations \cite{1,2,3,4,5,6}. The main goal of this study was to produce improved cross-section for the users, to calculate yield curves as well as to demonstrate the applicability of the produced radioisotopes for thin layer activation. The results may also contribute to the improvement of the theoretical model codes.
Here we present experimentally determined proton induced excitation functions on natural cobalt targets performed in different accelerator laboratories at 16, 37 and 70 MeV bombarding energies. For chosen radioisotopes yield curves were calculated from the measured cross-sections, as well as activity distribution (wear) curves are given for the most appropriate isotopes. The cross-section results are compared both with the available literature data and with the results of the most recent TALYS \cite{8} theoretical code taken from the TENDL-2011 library \cite{9}[9], as well as with the results of the EMPIRE 3.1 calculations \cite{10,11}.

\section{Experimental}
\label{sec:2}
For the experimental cross-section determination the stacked-foil technique was used. The measurements of the excitation functions were performed in three different experiments in two different accelerator laboratories. The first and second series of measurements were done in the Cyclotron Laboratory of the Free University of Brussels (VUB) at 37 and 16 MeV proton energies respectively. In the first experiment 50  m thick Co foils (Goodfellow) were stacked together with Cd (22.9  $\mu$m), Cu (12  $\mu$m) and Al (102  $\mu$m) foils as parallel measurement targets, monitor and energy degrader foils. The nominal beam intensity was 138 nA for 13.3 minutes at 37 MeV. In the second experiment at VUB also 50  m Co foils were stacked together with Cd (22.4  $\mu$m) and Cu (12  $\mu$m) foils without using any energy degraders. The irradiation was performed with 16 MeV proton beam for 25 minutes with 167 nA. The third series of measurements in Sendai, Japan (CYRIC) was performed at 70 MeV initial proton beam energy for 61 minutes with an average beam current of 58.2 nA. The stack was constructed from Co (50  $\mu$m), Er (25  $\mu$m) and In (50  $\mu$m) foils as parallel targets (Co, Er, In). Al (10 and 100  $\mu$m) foils were placed in the stack for monitoring the beam energy and intensity. After appropriate cooling time the stacks were disassembled and the foils were measured one by one by using HPGe detector based  $\gamma$-spectrometer systems in order to construct the excitation functions for the investigated nuclear reactions as well as for the monitor reactions. Several series of measurements have been performed, shortly after the end of bombardment (EOB) showing the $\gamma$-lines of shorter-lived isotopes (such as $^{56}$Ni) without larger contribution of the longer-lived peaks, and other 2 or 3 series later to see and measure the longer lived isotopes better. The nominal beam energies and intensities were corrected in each foil by using the $^{nat}$Al(p,x)$^{22,24}$Na monitor reactions. The excitation function of these reactions were re-measured in the whole investigated energy ranges and compared with the recommended values \cite{12}. The so calculated corrections were applied for each irradiation. 

\section{Theoretical calculations}
\label{sec:3}
The cross-sections of the investigated reactions were also compared with the cross-section data in the TENDL-2011 \cite{9} nuclear reaction data library constructed by the most recent version of the TALYS code as well as with the results of the EMPIRE 3.1 calculations. For the EMPIRE code, the optical potential parameters were taken from the Recommended Input Parameter Library RIPL-2 \cite{13}. The TENDL library was developed on the basis of the TALYS nuclear model code \cite{8} and the RIPL-2 \cite{13} parameters were used for the default calculations. The quality of the models in the TALYS code was improved starting with phenomenological default parameters for TALYS 1.0 and more microscopic options in TALYS 1.2 version \cite{14}. In TENDL 2011 significant improvement is seen for the predictive power for the activation cross-sections of proton induced reactions according to our experience. 
The TALYS and EMPIRE-3 codes calculate population of different low-lying levels and can also estimate the isomeric ratios for these levels. For all identified radioisotope the new EMPIRE-3.1 (Rivoli) was applied \cite{10,11}. From the point of view of charged particle induced reaction the new version of EMPIRE code contains the following new features: RIPL-3 library of input parameters; new version of Coupled Channel code ECIS-2006; coupled Channel code OPTMAN for soft-rotor calculations; parity dependent level densities; new parameterization of EGSM level densities; three additional ejectiles (d, t,$^3$He).

\section{Determination of the excitation functions}
\label{sec:4}
The activities of the irradiated target foils were measured without chemical separation by using high resolution  $\gamma$-ray spectrometry. The decay of the samples was followed by measuring each sample in several times (3-4 measurements), starting from a few hours after End of Bombardment (EOB) until weeks later. The cross-sections and their uncertainties were determined by using the well-known activation and the decay formulas \cite{15}. The decay data of the investigated isotopes and the Q-values of the contributing processes were taken from NUDAT \cite{16} and from the Q-value calculator of NNDC \cite{17} as shown in Table 1. The uncertainty of the measured values was estimated in a standard way \cite{18}: the independent relative errors of the linearly contributing processes (number of the bombarding particles (8\%), number of the target nuclei (5\%), decay data (3\%), detector efficiency (5\%) and peak area (1-7\%)) were summed quadratically and the square root of the sum was taken. The uncertainty of the bombarding energy was estimated to be 0.2-0.3 MeV by the first foil and increases linearly along the stack. 

\begin{table*}
\caption{\textbf{Nuclear data of the investigated reactions}}
\label{tab:1}       
\begin{tabular}{|p{0.5in}|p{0.5in}|p{0.7in}|p{0.5in}|p{0.5in}|p{1.1in}|p{0.9in}|} \hline 
\textbf{Nuclei} & \textbf{Half-life} & \textbf{Decay mode (\%)} & \textbf{E${}_{{\mathbf \gamma }}$\newline (keV)} & \textbf{I${}_{{\mathbf \gamma }}$\newline (\%)} & \textbf{Production route} & \textbf{Q-value (MeV)} \\ \hline 
${}^{51}$Cr & 27.7 d & EC(100) & 320.082 & 9.91 & ${}^{59}$Co(p,n2$\alpha$) & -16.41 \\ \hline 
${}^{52g}$Mn & 5.591 d & EC, ${\beta}^{+}$\newline (70.4, 29.6) & 744.233\newline 935.544\newline 1434.092 & 90.0\newline 94.5\newline 100 & ${}^{59}$Co(p,p3n$\alpha$)\newline ${}^{59}$Co(p,d2n$\alpha$)\newline  & -38.16\newline -35.94 \\ \hline 
${}^{54}$Mn & 312.3 d & EC(100)\newline ${\beta}^{-}$($<$2.9e-4) & 834.848\newline  & 99.976\newline  & ${}^{59}$Co(p,pn$\alpha$)\newline ${}^{59}$Co(p,d$\alpha$) & -17.17\newline -14.94 \\ \hline 
${}^{56}$Mn & 2.5785 h & ${\beta}^{-}$(100) & 846.771\newline 1238.232 & 98.9\newline 0.099 & ${}^{59}$Co(p,3pn)\newline ${}^{59}$Co(p,d2p) & -27.97\newline -25.74 \\ \hline 
${}^{55}$Co & 17.53 h & EC, ${\beta}^{+}$\newline (24, 76) & 477.2\newline 931.1\newline 1408.5 & 20.2\newline 75.0\newline 16.9 & ${}^{59}$Co(p,4np)\newline ${}^{59}$Co(p,5n)${}^{55}$Ni decay\newline  & -40.49\newline -49.96 \\ \hline 
${}^{56}$Co & 77.27 d & EC, ${\beta}^{+}$\newline (80.3, 19.7) & 846.771\newline 1238.232 & 100\newline 67.6 & ${}^{59}$Co(p,nt)\newline ${}^{59}$Co(p,4n)${}^{56}$Ni decay & -21.92\newline -33.32 \\ \hline 
${}^{57}$Co & 271.79 d & EC(100) & 122.0614\newline 136.4743 & 85.6\newline 10.68 & ${}^{59}$Co(p,2np)\newline ${}^{59}$Co(p,3n)${}^{57}$Ni decay & -19.03\newline -23.07 \\ \hline 
${}^{58}$${}^{g}$Co & 70.86 d & EC, ${\beta}^{+}$\newline (83.1, 14,9) & 810.775 & 99.0 & ${}^{59}$Co(p,np) & -10.45 \\ \hline 
${}^{56}$Ni & 6.075 d & EC(100) & 158.38\newline 749.95\newline 811.85 & 98.8\newline 49.5\newline 86.0 & ${}^{59}$Co(p,4n) & -33.32 \\ \hline 
${}^{57}$Ni & 35.6 h & EC, ${\beta}^{+}$\newline (56.4, 43.6) & 127.164\newline 1377.63 & 16.7\newline 81.7 & ${}^{59}$Co(p,3n) & -23.07 \\ \hline 
\end{tabular}
\begin{flushleft}
\footnotesize{When complex particles are emitted instead of individual protons and neutrons the Q-values have to be decreased by the respective binding energies of the compound particles: np-d, +2.2 MeV; 2np-t, +8.48 MeV; n2p-${}^{3}$He, +7.72 MeV; 2n2p-$\alpha$, +28.30 MeV}
\end{flushleft}
\end{table*}

\subsection{Measurement of the radioisotopes of Ni}
\label{sec:4.1}
The radioisotopes $^{56,57}$Ni having appropriate half-life and  $\gamma$-energies could be observed in our experiment. Nickel isotopes are produced by (p,xn) reactions from $^{59}$Co. According to the threshold energies of the reactions $^{54,55}$Ni could also be produced in our high energy experiment (series 3), but because of their short half-life we could not observe them in the measured spectra. Their production must be taken into account in the theoretical calculations as mother isotope of the produced longer lived Co isotopes.

\subsubsection{$^{59}$Co(p,3n)$^{57}$Ni reaction}
\label{sec:4.1.1}
The $^{57}$Ni radioisotope can only be produced through the above reaction and due to its threshold (Table 1) could only be detected in experiments 1. (37 MeV) and 3. (70 MeV). The results are presented in Fig. 1 together with the literature data and the results of the model calculations. Our data are in good agreement with the earlier results of Haasbroek \cite{19} and Michel \cite{20,21} and also with Sharp \cite{22} in the higher energy range. The results of other authors \cite{23,24,25} show only partial or no agreement with our new data. Both TALYS (TENDL 2011) and EMPIRE 3.1 give good estimation up to 30 MeV, but above this energy TALYS strongly underestimates the experimental results and also its prediction for the maximum is too low. EMPIRE 3.1 gives a good estimation for the excitation function maximum and only slightly underestimates above 60 MeV proton energy.

\begin{figure}
\includegraphics[width=0.55\textwidth]{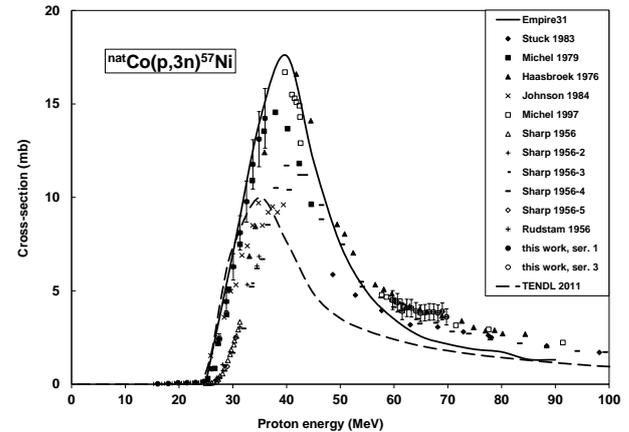}
\caption{Measured excitation function of the $^{nat}$Co(p,3n)$^{57}$Ni nuclear reaction compared with the literature and the results of the theoretical model codes}
\label{fig:1}       
\end{figure}

\subsubsection{$^{59}$Co(p,4n)$^{56}$Ni reaction}
\label{sec:4.1.2}
The $^{56}$Ni radioisotope can only be produced through the (p,4n) reaction and due to its threshold (Table 1) could only be detected in experiments 3. (70 MeV). The results are presented in Fig. 2 together with the literature data and the results of the model calculations. Our data are in god agreement with those of Haasbroek \cite{19} and Michel \cite{21} again. From the theoretical calculations the EMPIRE 3.1 gives better agreement again with our data in the higher energy region ($\>$60 MeV). Under 45 MeV both EMPIRE 3.1 and TENDL 2011 give very similar and acceptable results.

\begin{figure}
  \includegraphics[width=0.55\textwidth]{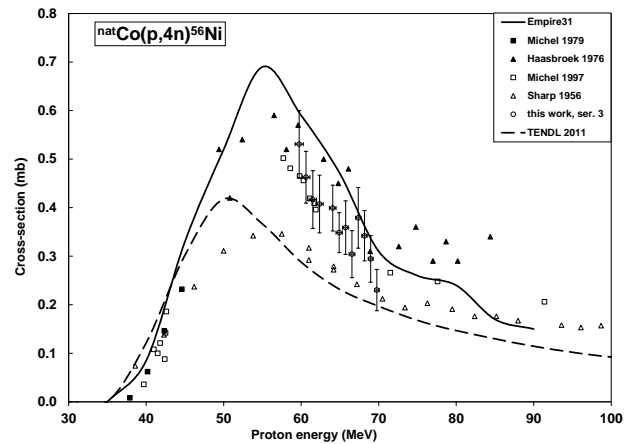}
\caption{Measured excitation function of the $^{nat}$Co(p,4n)$^{56}$Ni nuclear reaction compared with the literature and the results of the theoretical model codes}
\label{fig:2}       
\end{figure}

\subsection{Measurement of the radioisotopes of Co}
\label{sec:4.2}
The cobalt isotopes can be produced by (p,xnp) reaction from $^{59}$Co or by decay of the produced radioisotopes of nickel.

\subsubsection{$^{59}$Co(p,x)$^{58}$Co reaction}
\label{sec:4.2.1}
The $^{58}$Co radioisotope is produced with (p,np) reaction. It also has an excited state with 8.94 h half-life, but only with 24.9 keV  $\gamma$-energy, which was undetectable with our spectrometry system. The results for $^{58g}$Co, after the complete decay of the excited state, are presented in Fig. 3. We have now values from all series. Our new data are in good agreement with the previous results of Stuck \cite{24}, Haasbroek [19], Michel [21] over 55 MeV, and with Michel \cite{20,21} under 20 MeV. The data of other authors (\cite{26,27,28,29}) give slightly different results, while the data given by Sharp \cite{22} completely underestimate all other results. In this case the best theoretical estimation is given by TENDL 2011, while EMPIRE 3.1 underestimates above 35 MeV. 

\begin{figure}
  \includegraphics[width=0.55\textwidth]{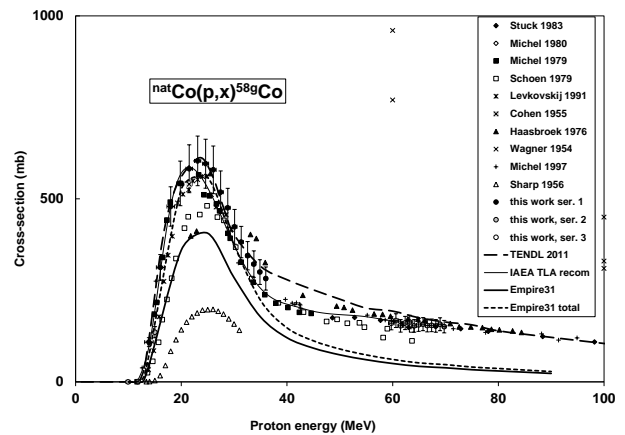}
\caption{Measured excitation function of the $^{nat}$Co(p,np)$^{58g}$Co nuclear reaction compared with the literature and the results of the theoretical model codes}
\label{fig:3}       
\end{figure}

\subsubsection{$^{59}$Co(p,x)$^{57}$Co (cum) reaction}
\label{sec:4.2.2}
The $^{57}$Co radioisotope is produced with (p,2np) reaction, as well as it comes from the decay of $^{57}$Ni. The measured data are presented in Fig. 4 together with the results of previous measurements as well as of the theoretical model calculations. Our new data are in good agreement with the previous results of Haasbroek \cite{19} and Michel \cite{21} above 50 MeV and with Michel \cite{20,21}, Levkovskij \cite{27} and Haasbroek under 40 MeV. The data of Stuck \cite{24} are lower than ours in the higher energy range, the Johnson's data \cite{23} decline significantly around the maximum while the previous results of Sharp \cite{22} seriously underestimate at the whole energy domain. The theoretical model calculations give similar results up the 50 MeV except the values of EMPIRE 3.1 are better around the maximum, while in the higher energy range above 50 MeV EMPIRE 3.1 strongly underestimates and TENDL 2011 slightly overestimates the experimental values. 

\begin{figure}
  \includegraphics[width=0.55\textwidth]{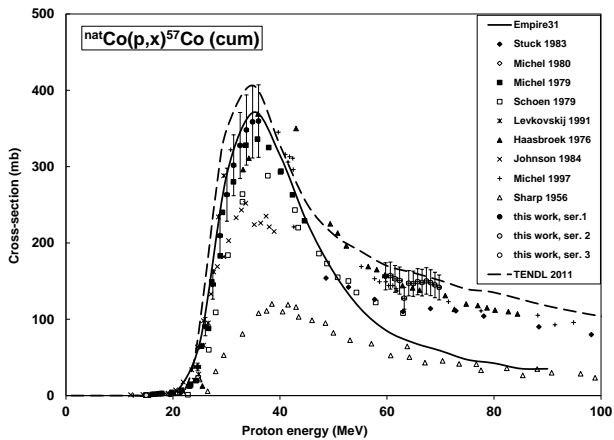}
\caption{Measured excitation function of the $^{nat}$Co(p,x)$^{57}$Co nuclear reaction compared with the literature and the results of the theoretical model codes}
\label{fig:4}       
\end{figure}

\subsubsection{$^{59}$Co(p,x)$^{56}$Co (cum) reaction}
\label{sec:4.2.3}
The $^{56}$Co radioisotope is produced with (p,3np) reaction, as well as it comes from the decay of $^{56}$Ni. The measured data are presented in Fig. 5 together with the results of previous measurements as well as of the theoretical model calculations. In this case we have only two series of measurements evaluated, because of the higher threshold of this reaction there was no $^{56}$Co detected from the 16 MeV experiment. According to Fig. 5 our data agree acceptably with those of Schoen \cite{26} and Sharp \cite{22} in the higher energy region and with the data of Johnson \cite{23}, Michel \cite{20} under 35 MeV. The data of Haasbroek \cite{19} overestimate, while those of Stueck [24] underestimate our results above 55 MeV. Wagner \cite{29} gave very scattered values, while the results of Sharp \cite{22} are too low again. TENDL 2011 and EMPIRE 3.1 give similar results up to 70 MeV again, while above this energy EMPIRE 3.1 declines significantly.	

\begin{figure}
  \includegraphics[width=0.55\textwidth]{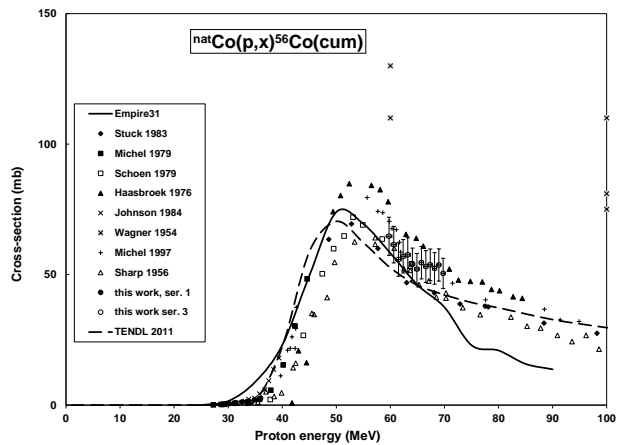}
\caption{Measured excitation function of the $^{nat}$Co(p,x)$^{56}$Co nuclear reaction compared with the literature and the results of the theoretical model codes}
\label{fig:5}       
\end{figure}

\subsubsection{$^{59}$Co(p,x)$^{55}$Co reaction}
\label{sec:4.2.4}
The $^{55}$Co radioisotope is produced with (p,4np) reaction, as well as it comes from the decay of $^{55}$Ni (threshold is 51 MeV). The measured data are presented in Fig. 6 together with the results of previous measurements as well as of the theoretical model calculations. Because of the higher threshold we have results only from the 70 MeV experiment (series 3.). Our new results now are slightly lower than the other experimental results. TENDL 2011 and EMPIRE 3.1 are similar at lower energies. EMPIRE 3.1 follows better the experimental trend up to 70 MeV but shows strange behavior (unexpected local maximum and minimum) above it. 

\begin{figure}
  \includegraphics[width=0.55\textwidth]{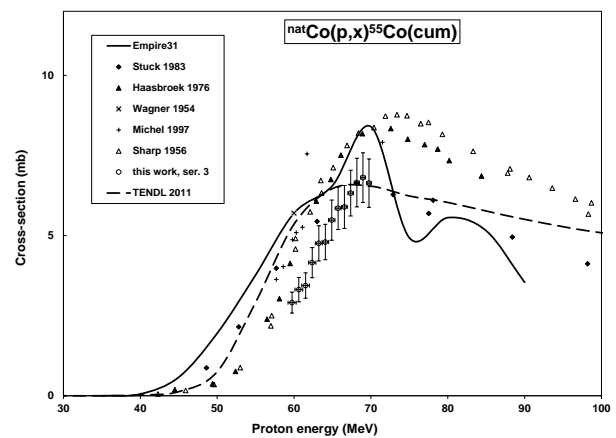}
\caption{Measured excitation function of the $^{nat}$Co(p,x)$^{55}$Co nuclear reaction compared with the literature and the results of the theoretical model codes}
\label{fig:6}       
\end{figure}

\subsection{Measurement of the radioisotopes of Mn}
\label{sec:4.3}
The manganese isotopes can be produced by (p,xnyp) reaction from $^{59}$Co, or a group of neutrons and protons can also form complex ejectiles, which is energetically more favorable. 

\subsubsection{$^{59}$Co(p,x)$^{56}$Mn reaction}
\label{sec:4.3.1}
The threshold of the $^{59}$Co(p,x)$^{56}$Mn reactions is above 40 MeV (see Table 1), coupled with relative short half-life (2.58 h). The TENDL 2011 library proposes quite low cross-section, unfortunately we could not even detect it in the 70 MeV experiment.

\subsubsection{$^{59}$Co(p,x)$^{54}$Mn reaction}
\label{sec:4.3.2}
The $^{54}$Mn radioisotope is produced with (p,3n3p) reaction. From the particles 3n3p complex particles ($\alpha$, d, $^3$He) can be formed with lower threshold energies (see Table 1). The measured data are presented in Fig. 7 together with the results of previous measurements as well as of the theoretical model calculations. Our new data are in god agreement with those of Stueck \cite{24}, Michel \cite{21} and Sharp \cite{22} above 55 MeV and with Michel \cite{20} under 35 MeV. The EMPIRE 3.1 code gives a good approximation up to 35 MeV, but strongly overestimates above it. The prediction of TENDL 2011 is too low under 55 MeV and it tends to give acceptable results above 70 MeV.

\begin{figure}
  \includegraphics[width=0.55\textwidth]{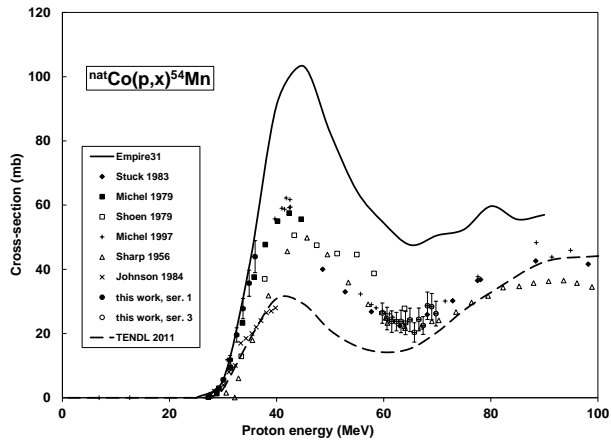}
\caption{Measured excitation function of the $^{nat}$Co(p,x)$^{54}$Mn nuclear reaction compared with the literature and the results of the theoretical model codes}
\label{fig:7}       
\end{figure}

\subsubsection{$^{59}$Co(p,x)$^{52g}$Mn reaction}
\label{sec:4.3.3}
The $^{52g}$Mn radioisotope is produced with reactions involving complex ejectiles (see Table 1). The short-lived isomeric state (T$_{1/2}$=21 min) could not be detected in our measurements. It has 1.75\% internal conversion to the ground state, so our results contain this contribution. The threshold is relatively high so we have only results from the 70 MeV experiment. The measured data are presented in Fig. 8 together with the results of previous measurements as well as of the theoretical model calculations. Having relatively few data sets for comparison we can only conclude that our data agree with those of Wagner \cite{29} at 60 MeV and slightly lower than others. The results of Haller \cite{30} are completely out of the trend of the experimental data and theoretical predictions. Both TENDL 2011 and EMPIRE 3.1 describe the trend relatively good up to 60 MeV, but above this energy EMPIRE 3.1 gives a better approximation.

\begin{figure}
  \includegraphics[width=0.5\textwidth]{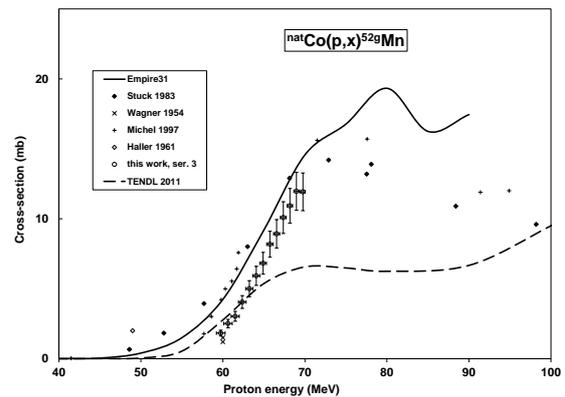}
\caption{Measured excitation function of the $^{nat}$Co(p,x)$^{52g}$Mn nuclear reaction compared with the literature and the results of the theoretical model codes}
\label{fig:8}       
\end{figure}

\subsection{Measurement of the radioisotopes of Cr}
\label{sec:4.4}
The only chromium radioisotope to be produced by cobalt irradiation is the $^{51}$Cr thanks to the formation of complex ejectiles (p,n2$\alpha$).

\subsubsection{$^{59}$Co(p,x)$^{51}$Cr reaction}
\label{sec:4.4.1}
The radioisotope $^{51}$Cr (T$_{1/2}$= 27.7 d) was easily measurable in our high energy measurement. The measured data are presented in Fig. 9 together with the results of previous measurements as well as of the theoretical model calculations. Our new results are in good agreement with the data of Stueck \cite{24} in the measured energy range. The EMPIRE 3.1 follows better the trend but overestimates while TENDL 2011 underestimates the experimental values. 

\begin{figure}
  \includegraphics[width=0.5\textwidth]{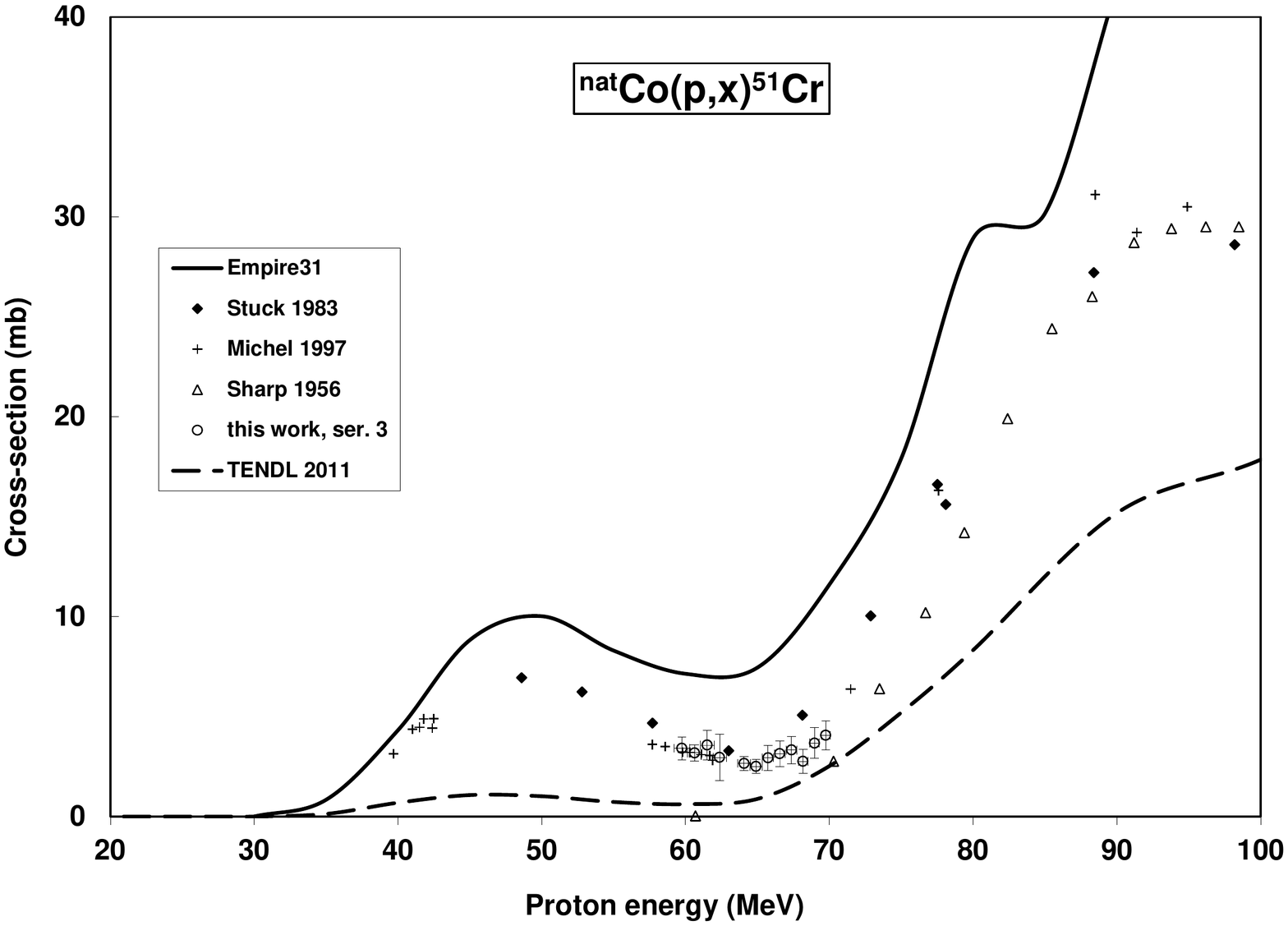}
\caption{Measured excitation function of the $^{nat}$Co(p,x)$^{51}$Cr nuclear reaction compared with the literature and the results of the theoretical model codes}
\label{fig:9}       
\end{figure}

\begin{table*}
\tiny
\caption{\textbf{Measured cross-section values}}
\label{tab:2}       
\begin{tabular}{|c|c|p{0.25 in}|p{0.2 in}|p{0.25 in}|p{0.2 in}|p{0.25 in}|p{0.2 in}|p{0.25 in}|p{0.2 in}|p{0.25 in}|p{0.2 in}|p{0.25 in}|p{0.2 in}|p{0.25 in}|p{0.15 in}|p{0.25 in}|p{0.15 in}|p{0.25 in}|p{0.15 in}|} \hline 
\multicolumn{2}{|c|}{\textbf{Bombarding}} & \multicolumn{2}{|c|}{\textbf{${}^{57}$Ni}} & \multicolumn{2}{|c|}{\textbf{${}^{56}$Ni}} & \multicolumn{2}{|c|}{\textbf{${}^{58g}$Co}} & \multicolumn{2}{|c|}{\textbf{${}^{57}$Co}} & \multicolumn{2}{|c|}{\textbf{${}^{56}$Co}} & \multicolumn{2}{|c|}{\textbf{${}^{55}$Co}} & \multicolumn{2}{|c|}{\textbf{${}^{54}$Mn}} & \multicolumn{2}{|c|}{\textbf{${}^{52g}$Mn}} & \multicolumn{2}{|c|}{\textbf{${}^{51}$Cr}} \\ \hline 
E & $\pm$dE & $\sigma$ & $\pm$d$\sigma$ & $\sigma$ & $\pm$d$\sigma$ & $\sigma$ & $\pm$d$\sigma$ & $\sigma$ & $\pm$d$\sigma$ & $\sigma$ & $\pm$d$\sigma$ & $\sigma$ & $\pm$d$\sigma$ & $\sigma$ & $\pm$d$\sigma$ & $\sigma$ & $\pm$d$\sigma$ & $\sigma$ & $\pm$d$\sigma$ \\ \hline 
\multicolumn{2}{|c|}{MeV} & \multicolumn{2}{|c|}{mb} & \multicolumn{2}{|c|}{mb} & \multicolumn{2}{|c|}{mb} & \multicolumn{2}{|c|}{mb} & \multicolumn{2}{|c|}{mb} & \multicolumn{2}{|c|}{mb} & \multicolumn{2}{|c|}{mb} & \multicolumn{2}{|c|}{mb} & \multicolumn{2}{|c|}{mb} \\ \hline 
9.9 & 0.4 &  & ~ &  & ~ & 0.03 & 0.01 &  & ~ &  & ~ &  & ~ &  & ~ &  & ~ &  & ~ \\ \hline 
11.8 & 0.3 &  & ~ &  & ~ & 0.65 & 0.08 &  & ~ &  & ~ &  & ~ &  & ~ &  & ~ &  & ~ \\ \hline 
13.5 & 0.3 &  & ~ &  & ~ & 40.9 & 4.6 &  & ~ &  & ~ &  & ~ &  & ~ &  & ~ &  & ~ \\ \hline 
13.9 & 0.8 &  & ~ &  & ~ & 108.5 & 12.2 &  & ~ &  & ~ &  & ~ &  & ~ &  & ~ &  & ~ \\ \hline 
15.0 & 0.3 &  & ~ &  & ~ & 135.6 & 15.2 & 0.38 & 0.08 &  & ~ &  & ~ &  & ~ &  & ~ &  & ~ \\ \hline 
16.0 & 0.8 & 0.02 & 0.01 &  & ~ & 312.8 & 35.1 & 1.3 & 0.2 &  & ~ &  & ~ &  & ~ &  & ~ &  & ~ \\ \hline 
18.0 & 0.7 & 0.04 & 0.01 &  & ~ & 479.1 & 53.8 & 2.6 & 0.4 &  & ~ &  & ~ &  & ~ &  & ~ &  & ~ \\ \hline 
19.8 & 0.7 & 0.08 & 0.01 &  & ~ & 542.3 & 60.9 & 3.1 & 0.5 &  & ~ &  & ~ &  & ~ &  & ~ &  & ~ \\ \hline 
21.5 & 0.6 & 0.08 & 0.01 &  & ~ & 582.2 & 65.4 & 4.4 & 0.6 &  & ~ &  & ~ &  & ~ &  & ~ &  & ~ \\ \hline 
23.1 & 0.6 & 0.09 & 0.01 &  & ~ & 603.7 & 67.8 & 12.3 & 1.7 &  & ~ &  & ~ &  & ~ &  & ~ &  & ~ \\ \hline 
24.6 & 0.6 & 0.15 & 0.02 &  & ~ & 596.1 & 66.9 & 37.5 & 5.0 &  & ~ &  & ~ &  & ~ &  & ~ &  & ~ \\ \hline 
26.1 & 0.5 & 0.82 & 0.09 &  & ~ & 579.4 & 65.0 & 90.3 & 11.9 &  & ~ &  & ~ &  & ~ &  & ~ &  & ~ \\ \hline 
27.4 & 0.5 & 2.4 & 0.3 &  & ~ & 518.2 & 58.2 & 145.6 & 19.2 & 0.13 & 0.04 &  & ~ & 0.27 & 0.14 &  & ~ &  & ~ \\ \hline 
28.8 & 0.5 & 4.4 & 0.5 &  & ~ & 475.5 & 53.4 & 209.5 & 27.7 & 0.31 & 0.07 &  & ~ & 2.12 & 0.3 &  & ~ &  & ~ \\ \hline 
30.1 & 0.4 & 6.3 & 0.7 &  & ~ & 423.7 & 47.6 & 263.2 & 34.8 & 0.51 & 0.08 &  & ~ & 5.6 & 0.7 &  & ~ &  & ~ \\ \hline 
31.3 & 0.4 & 8.1 & 0.9 &  & ~ & 382.0 & 42.9 & 301.8 & 39.9 & 0.87 & 0.11 &  & ~ & 11.8 & 1.3 &  & ~ &  & ~ \\ \hline 
32.5 & 0.4 & 9.8 & 1.1 &  & ~ & 344.8 & 38.7 & 327.7 & 43.3 & 1.1 & 0.2 &  & ~ & 19.6 & 2.2 &  & ~ &  & ~ \\ \hline 
33.7 & 0.4 & 11.8 & 1.3 &  & ~ & 321.7 & 36.1 & 348.0 & 46.0 & 1.3 & 0.2 &  & ~ & 27.8 & 3.2 &  & ~ &  & ~ \\ \hline 
34.9 & 0.3 & 13.1 & 1.5 &  & ~ & 300.0 & 33.7 & 358.7 & 47.4 & 2.0 & 0.3 &  & ~ & 35.7 & 4.1 &  & ~ &  & ~ \\ \hline 
36.0 & 0.3 & 14.2 & 1.6 &  & ~ & 282.3 & 31.7 & 359.7 & 47.5 & 2.6 & 0.3 &  & ~ & 43.9 & 4.9 &  & ~ &  & ~ \\ \hline 
59.7 & 0.5 & 4.5 & 0.5 & 0.53 & 0.07 & 160.2 & 18.0 & 156.3 & 17.8 & 64.7 & 7.3 & 2.9 & 0.3 & 26.4 & 3.2 & 1.8 & 0.2 & 3.4 & 0.6 \\ \hline 
60.6 & 0.5 & 4.7 & 0.5 & 0.46 & 0.05 & 167.4 & 18.8 & 157.2 & 17.7 & 61.5 & 7.0 & 3.3 & 0.4 & 24.7 & 3.5 & 2.5 & 0.3 & 3.2 & 0.4 \\ \hline 
61.5 & 0.5 & 3.9 & 0.5 & 0.42 & 0.06 & 153.0 & 17.2 & 152.9 & 17.6 & 55.8 & 6.4 & 3.4 & 0.4 & 23.7 & 3.4 & 3.0 & 0.4 & 3.6 & 0.7 \\ \hline 
62.4 & 0.5 & 4.1 & 0.5 & 0.41 & 0.06 & 160.9 & 18.1 & 150.7 & 17.8 & 57.0 & 6.4 & 4.2 & 0.5 & 23.7 & 2.9 & 4.1 & 0.5 & 2.9 & 1.1 \\ \hline 
63.2 & 0.5 & 4.1 & 0.5 & 0.00 & 0.00 & 163.5 & 18.4 & 127.5 & 14.3 & 57.6 & 6.6 & 4.8 & 0.6 & 23.8 & 3.5 & 5.0 & 0.6 &  &  \\ \hline 
64.1 & 0.4 & 3.9 & 0.5 & 0.40 & 0.05 & 152.9 & 17.2 & 147.2 & 16.6 & 54.0 & 6.2 & 4.8 & 0.6 & 23.3 & 3.6 & 5.9 & 0.7 & 2.7 & 0.4 \\ \hline 
64.9 & 0.4 & 3.8 & 0.4 & 0.35 & 0.04 & 155.6 & 17.5 & 146.6 & 16.5 & 52.1 & 6.0 & 5.5 & 0.6 & 24.3 & 3.7 & 6.8 & 0.8 & 2.5 & 0.3 \\ \hline 
65.7 & 0.4 & 3.8 & 0.4 & 0.36 & 0.06 & 157.6 & 17.7 & 149.3 & 17.1 & 54.6 & 6.2 & 5.9 & 0.7 & 20.4 & 3.0 & 8.2 & 0.9 & 2.9 & 0.6 \\ \hline 
66.6 & 0.4 & 3.9 & 0.5 & 0.30 & 0.05 & 153.9 & 17.3 & 148.0 & 17.4 & 53.1 & 6.1 & 5.9 & 0.7 & 24.4 & 3.6 & 8.9 & 1.0 & 3.1 & 0.6 \\ \hline 
67.4 & 0.4 & 3.9 & 0.4 & 0.38 & 0.06 & 154.4 & 17.3 & 150.3 & 17.3 & 53.8 & 6.1 & 6.3 & 0.7 & 22.5 & 2.8 & 10.1 & 1.1 & 3.3 & 0.7 \\ \hline 
68.2 & 0.3 & 3.8 & 0.4 & 0.34 & 0.05 & 152.9 & 17.2 & 147.9 & 16.9 & 52.4 & 6.1 & 6.7 & 0.8 & 28.7 & 4.3 & 10.9 & 1.2 & 2.8 & 0.6 \\ \hline 
\end{tabular}
\end{table*}

\section{Physical yield}
\label{sec:5}
From the point of view of radioisotope production the production yield versus energy curves are even more important than the excitation functions, because the given information is more closed to the applications. Searching the literature we have found experimental or calculated yields from the following authors: Dmitriev 1981 \cite{31}, Haasbroek \cite{19} and Abe \cite{32}. Our calculated data are in good agreement with the data points of Dmitriev for $^{57}$Co and $^{58g}$Co (see Fig. 10). The measured curves of Haasbroek are also in good agreement with ours in the case of $^{56}$Ni, $^{55}$Co and $^{56}$Co over 50 MeV, $^{57}$Co over 40 MeV, $^{58g}$Co over 25 MeV, under these energies there are only small declinations. In the case of $^{57}$Ni we are slightly above the Haasbroek values. 

\begin{figure}
  \includegraphics[width=0.5\textwidth]{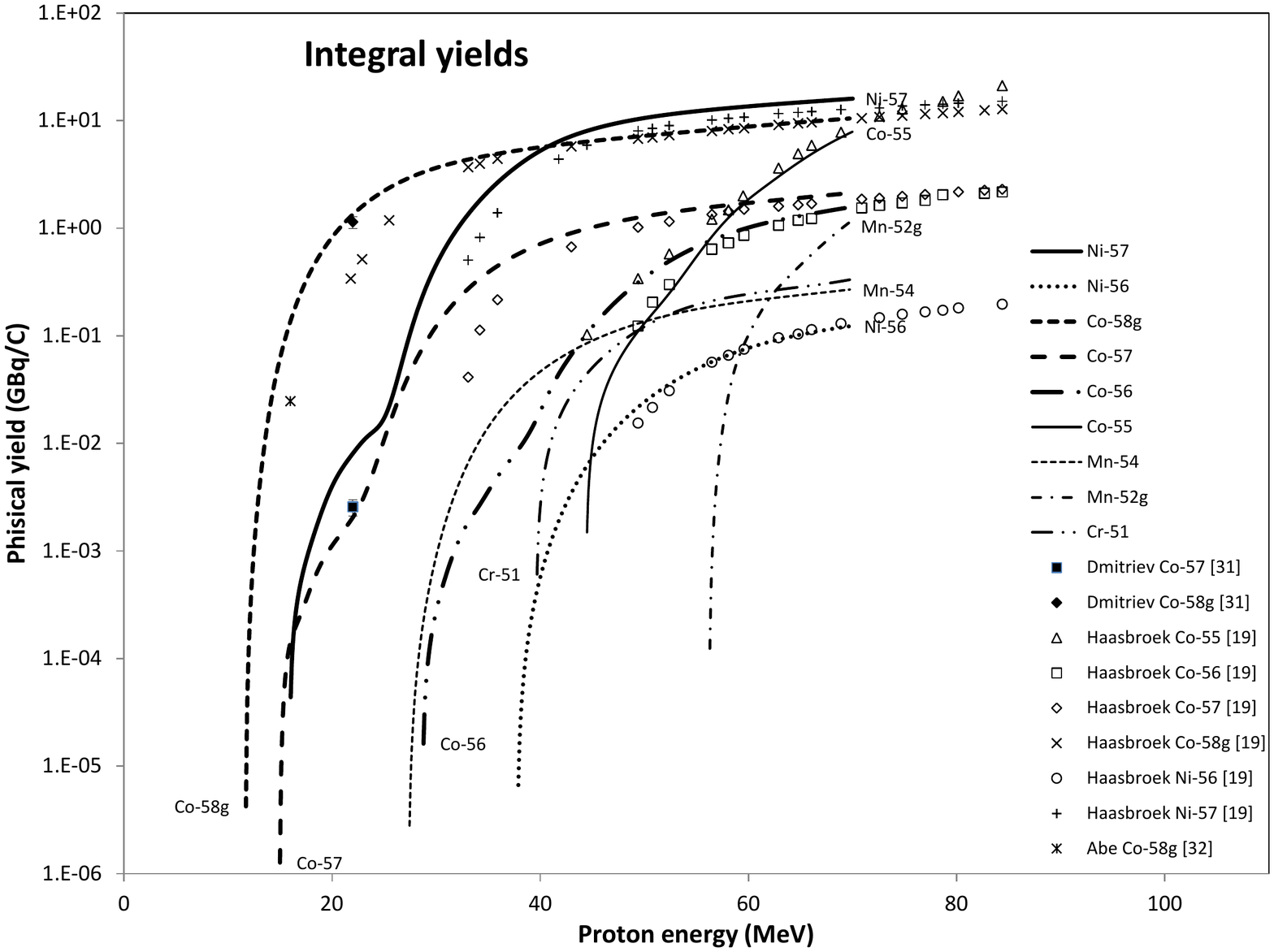}
\caption{Physical yield curves calculated from the measured cross-sections: for fitting of the excitation functions we have used additionally to the results of this work the following data: $^{57}$Ni: Michel 1997; $^{56}$Ni: Michel 1979 + Michel 1997; $^{57}$Co: Michel 1997; $^{56}$Co: Michel 1997; $^{55}$Co: Haasbroek 1976; $^{54}$Mn: Michel 1997; $^{52g}$Mn: Michel 1997; $^{51}$Cr: Stueck 1983+Michel 1997}
\label{fig:10}       
\end{figure}

\section{Thin layer activation}
\label{sec:6}
A common use of radioisotope tracers is the study of wear, corrosion and erosion behaviors of different materials by using thin layer activation (TLA). Among the investigate isotopes in Table 1 we have selected those, which fulfill the following requirements for wear measurements:
\begin{itemize}
\item Long enough half-life to perform the measurements ($^{51}$Cr(27.7 d), $^{52g}$Mn(5.591 d), $^{54}$Mn(312.3 d), $^{56}$Co(77.26 d), $^{57}$Co(271.79 d), $^{58g}$Co(70.86 d), $^{56}$Ni(6.076 d))
\item High enough cross-section/yield to produce considerable activity within reasonable irradiation time 
\item Strong  $\gamma$-line(s) with suitable energy
\end{itemize}
The first criterion is fulfilled for the above listed isotopes. If we consider only irradiations below 35 MeV, then only $^{58g}$Co and $^{57}$Co remain. The TLA irradiation can be performed in two ways: 1. the so called "homogeneous" activity distribution means that the activity/layer thickness is the same in the top several micrometers of the activated surface. This can be achieved by choosing irradiation energy around the excitation function maximum (if any). These bombarding energies are 36 MeV and 22 MeV for $^{57}$Co and $^{58g}$Co respectively. 2. The "linear" activity distribution is produced if the excitation curve has no maximum within the available energy range, or because of the lower requested penetration a lower bombarding energy is selected. In Fig. 11 the two cases for $^{58g}$Co are demonstrated. The optimum irradiation energy for homogeneous activity distribution is 22.4 MeV (solid curve), which results in constant activity (within 1\%) for 121 $\mu$m depth. The scattered line represents the linear activity distribution, for which 15 MeV bombarding energy was chosen, and the linearity is valid to 42 $\mu$m depth. Similar calculations for $^{57}$Co are also shown in Fig 11. In this case the optimum energy for homogeneous activity distribution is 31.3 MeV, while the "linear" curve is produced by 25 MeV irradiation. The homogeneity and linearity ranges are 167 and 162 $\mu$m respectively. The 10 day cooling time applied in all cases is convenient period for sample measurement, preparation and possible delivery to other site. From Fig. 11 it is seen that in the case of $^{58g}$Co higher specific activities can be produced.

\begin{figure}
  \includegraphics[width=0.5\textwidth]{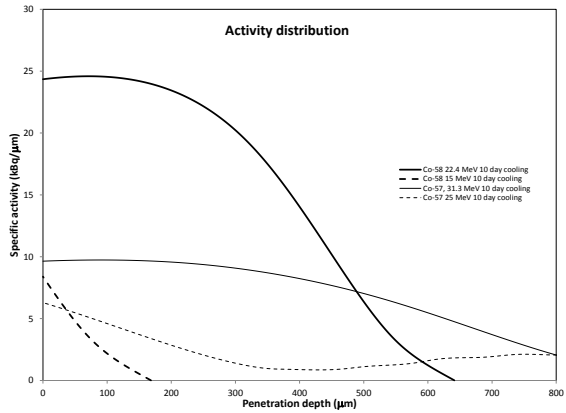}
\caption{11.	Wear curves of different irradiations for $^{58g}$Co and $^{57}$Co: irradiation time = 1 hour, beam current = 2 $\mu$A. 10 day cooling time was applied in all cases.}
\label{fig:11}       
\end{figure}

\section{Summary and conclusion}
\label{sec:7}
Excitation functions of radioisotopes produced from natural cobalt ($^{59}$Co) have been measured in different accelerator laboratories by using the method of complete re-measurement of well-known monitor reactions, and compared with the existing literature data as well as with the results of different model calculations (EMPIRE-3, TALYS(TENDL 2011). Our new measurements showed acceptable agreement with the existing experimental data if available, and helped to decide if contradicting literature results exist. The results of the model calculations showed an improvement compared with the earlier versions, but there are still discrepancies especially in the higher energy range. Altogether, the both theoretical models in scope describe acceptable well the shape of the excitation functions up to 40 MeV in most cases, and also give estimation for the maximum and the maximum value. Our new data can be compared with the new versions of the theoretical model codes to elaborate an improved version of them. Yield calculations have also been performed and compared with the literature data if available. Our new results are in good agreement with the available experimental data. A graphical comparison is also given to help the users from medical or industrial field to choose the appropriate nuclear reaction. For the most suitable radioisotopes for TLA wear curves are also calculated and proved their applicability for industrial use.


%
%

\begin{acknowledgements}
This study was partly performed in the frame of the MTA-JSPS and MTA-FWO (Vlaanderen) collaboration programs. The authors thank the different research projects and their respective institutions for the practical help and providing the use of the facilities for this study.
\end{acknowledgements}

\bibliographystyle{spphys}       
\bibliography{Co+p}   

\begin{thebibliography}{10}
\providecommand{\url}[1]{{#1}}
\providecommand{\urlprefix}{URL }
\expandafter\ifx\csname urlstyle\endcsname\relax
  \providecommand{\doi}[1]{DOI \discretionary{}{}{}#1}\else
  \providecommand{\doi}{DOI \discretionary{}{}{}\begingroup
  \urlstyle{rm}\Url}\fi

\bibitem{7}
S.~Spellerberg, P.~Reimer, G.~Blessing, H.H. Coenen, S.M. Qaim, Applied
  Radiation and Isotopes \textbf{49}(12), 1519 (1998)

\bibitem{1}
F.~T\'ark\'anyi, F.~Ditr\'oi, F.~Szelecs\'enyi, M.~Sonck, A.~Hermanne, Nuclear
  Instruments \& Methods in Physics Research Section B-Beam Interactions with
  Materials and Atoms \textbf{198}(1-2), 11 (2002)

\bibitem{2}
F.~T\'ark\'anyi, F.~Ditr\'oi, S.~Tak\'acs, J.~Csikai, A.~Hermanne, M.S. Uddin,
  M.~Hagiwara, M.~Baba, Y.N. Shubin, A.I. Dityuk, Nuclear Instruments \&
  Methods in Physics Research Section B-Beam Interactions with Materials and
  Atoms \textbf{226}(4), 473 (2004)

\bibitem{3}
M.S. Uddin, M.~Hagiwara, M.~Baba, F.~T\'ark\'anyi, F.~Ditr\'oi, Applied
  Radiation and Isotopes \textbf{63}(3), 367 (2005)

\bibitem{4}
A.~Hermanne, F.~T\'ark\'anyi, F.~Ditr\'oi, S.~Tak\'acs, R.A. Rebeles, M.S.
  Uddin, M.~Hagiwara, M.~Baba, Y.~Shubin, S.F. Kovalev, Nuclear Instruments \&
  Methods in Physics Research Section B-Beam Interactions with Materials and
  Atoms \textbf{247}(2), 180 (2006)

\bibitem{5}
F.~Ditr\'oi, F.~T\'ark\'anyi, S.~Tak\'acs, A.~Hermanne, M.~Baba, A.V. Ignatyuk,
  Nuclear Instruments \& Methods in Physics Research Section B-Beam
  Interactions with Materials and Atoms \textbf{268}(17-18), 2571 (2010)

\bibitem{6}
F.~Ditr\'oi, S.~Tak\'acs, F.~T\'ark\'anyi, R.W. Smith, M.~Baba, Journal of the
  Korean Physical Society \textbf{59}(2), 1697 (2011)

\bibitem{8}
A.J. Koning, S.~Hilaire, M.C. Duijvestijn.
\newblock Talys-1.0 (2007)

\bibitem{9}
A.J. Koning, D.~Rochman.
\newblock Talys-based evaluated nuclear data library version 4 (2011)

\bibitem{10}
M.~Herman, R.~Capote, B.V. Carlson, P.~Oblozinsky, M.~Sin, A.~Trkov, H.~Wienke,
  V.~Zerkin, Nuclear Data Sheets \textbf{108}(12), 2655 (2007)

\bibitem{11}
M.~Herman, R.~Capote, M.~Sin, A.~Trkov, B.~Carlson, P.~Oblozinsky, C.~Mattoon,
  H.~Wienke, S.~Hoblit, Y.S. Cho, V.~Plujko, V.~Zerkin.
\newblock Nuclear reaction model code empire-3.1 (rivoli): (2012)

\bibitem{12}
F.~T\'ark\'anyi, S.~Tak\'acs, K.~Gul, A.~Hermanne, M.G. Mustafa, M.~Nortier,
  P.~Oblozinsky, S.M. Qaim, B.~Scholten, Y.N. Shubin, Z.~Youxiang, Beam monitor
  reactions (chapter 4). charged particle cross-section database for medical
  radioisotope production: diagnostic radioisotopes and monitor reactions.
\newblock Tech. rep., IAEA (2001)

\bibitem{13}
T.~Belgya, O.~Bersillon, R.~Capote, T.~Fukahori, G.~Zhigang, S.~Goriely,
  M.~Herman, A.V. Ignatyuk, S.~Kailas, A.~Koning, P.~Oblozinsky, V.~Plujko,
  P.~Young, \emph{Handbook for calculations of nuclear reaction data: Reference
  Input Parameter Library. http://www-nds.iaea.org/RIPL-2/} (IAEA, Vienna,
  2005)

\bibitem{14}
S.~Hilaire, A.J. Koning, S.~Goriely, EPJ Web of Conferences \textbf{8}, 02004
  (2010)

\bibitem{15}
G.~Deconninck, \emph{Introduction to radioanalytical physics}.
\newblock Nuclear methods monographs (Elsevier Scientific Pub. Co. ;
  distribution for the U.S.A. and Canada, Elsevier/North-Holland, Amsterdam ;
  New York New York, 1978)

\bibitem{16}
NuDat.
\newblock Nudat 2.5 database http://www.nndc.bnl.gov/nudat2/ (2011)

\bibitem{17}
B.~Pritychenko, A.~Sonzogni.
\newblock Q-value calculator (2003)

\bibitem{18}
I.B. of-Weights-and Measures, \emph{Guide to the expression of uncertainty in
  measurement}, 1st edn. (International Organization for Standardization,
  Genève, Switzerland, 1993)

\bibitem{19}
F.J. Haasbroek, J.~Steyn, R.D. Neirinckx, G.F. Burdzik, M.~Cogneau, P.~Wanet,
  International Journal of Applied Radiation and Isotopes \textbf{28}(5), 533
  (1977)

\bibitem{20}
R.~Michel, G.~Brinkmann, H.~Weigel, W.~Herr, Nuclear Physics A \textbf{322}(1),
  40 (1979)

\bibitem{21}
R.~Michel, R.~Bodemann, H.~Busemann, R.~Daunke, M.~Gloris, H.J. Lange, B.~Klug,
  A.~Krins, I.~Leya, M.~Lupke, S.~Neumann, H.~Reinhardt, M.~SchnatzButtgen,
  U.~Herpers, T.~Schiekel, F.~Sudbrock, B.~Holmqvist, H.~Conde, P.~Malmborg,
  M.~Suter, B.~DittrichHannen, P.W. Kubik, H.A. Synal, D.~Filges, Nuclear
  Instruments \& Methods in Physics Research Section B-Beam Interactions with
  Materials and Atoms \textbf{129}(2), 153 (1997)

\bibitem{22}
R.A. Sharp, R.M. Diamond, G.~Wilkinson, Physical Review \textbf{101}, 1493
  (1956)

\bibitem{23}
P.C. Johnson, M.C. Lagunassolar, M.J. Avila, International Journal of Applied
  Radiation and Isotopes \textbf{35}(5), 371 (1984)

\bibitem{24}
R.~Stueck, Proton induced reactions on ti, v, mn, fe, co and ni. measurement
  and hybrid model analysis of integral excitation functions and their
  application in model calculation for the production of cosmogenic nuclides.
\newblock Ph.D. thesis (1983)

\bibitem{25}
G.~Rudstam, Phd thesis.
\newblock Ph.D. thesis (1956)

\bibitem{26}
N.C. Schoen, G.~Orlov, R.J. Mcdonald, Physical Review C \textbf{20}(1), 88
  (1979)

\bibitem{27}
V.N. Levkovskii, \emph{The cross-sections of activation of nuclides of
  middle-range mass (A=40-100) by protons and alpha particles of middle range
  energies (E=10-50 MeV)} (Inter-Vesy, Moscow, 1991)

\bibitem{28}
B.L. Cohen, E.~Newman, T.H. Handley, Physical Review \textbf{99}(3), 723 (1955)

\bibitem{29}
G.D. Wagner, E.O. Wiig, Physical Review \textbf{96}(4), 1100 (1954)

\bibitem{30}
I.~Haller, G.~Rudstam, Inorganic \& Nuclear Chemistry \textbf{19}, 1 (1961)

\bibitem{31}
P.P. Dmitriev, G.A. Molin, Vop. At. Nauki i Tekhn., Ser.Yadernye Konstanty
  \textbf{44}(5), 43 (1981)

\bibitem{32}
K.~Abe, A.~Iizuka, A.~Hasegawa, S.~Morozumi, Journal of Nuclear Materials
  \textbf{123}(1-3), 972 (1984)

\end{thebibliography}

%
%

\end{document}